\documentclass[aps,pra,twocolumn,notitlepage,superscriptaddress,showpacs,nofootinbib]{revtex4-2}
\usepackage{amsmath}
\usepackage{amssymb}
\usepackage{amsfonts}
\usepackage{enumerate}
\usepackage{mathrsfs}
\usepackage{booktabs}
\usepackage{graphicx}
\usepackage{float}     
\usepackage{array}     
\usepackage{epstopdf}
\usepackage{enumerate}
\usepackage{changepage}
\usepackage{bm}
\usepackage{multirow}
\usepackage{lipsum}
\usepackage{booktabs}
\usepackage[ruled,vlined]{algorithm2e}  
\usepackage{mathtools}  
\usepackage{tikz}
\usepackage{adjustbox} 

\newtheorem{theorem}{Theorem}
\newtheorem{definition}{Definition}
\newtheorem{observation}{Observation}
\newtheorem{lemma}{Lemma}
\newtheorem{proposition}{Proposition}
\newtheorem{conjecture}{Conjecture}
\newtheorem{example}{Example}
\newtheorem{corollary}{Corollary}

\def\bcj{\begin{conjecture}}
	\def\ecj{\end{conjecture}}
\def\bcr{\begin{corollary}}
	\def\ecr{\end{corollary}}
\def\bd{\begin{definition}}
	\def\ed{\end{definition}}
\def\bea{\begin{eqnarray}}
	\def\eea{\end{eqnarray}}
\def\bem{\begin{enumerate}}
	\def\eem{\end{enumerate}}
\def\bex{\begin{example}}
	\def\eex{\end{example}}
\def\bim{\begin{itemize}}
	\def\eim{\end{itemize}}
\def\bl{\begin{lemma}}
	\def\el{\end{lemma}}
\def\bma{\begin{bmatrix}}
	\def\ema{\end{bmatrix}}
\def\bpf{\begin{proof}}
	\def\epf{\end{proof}}
\def\bpp{\begin{proposition}}
	\def\epp{\end{proposition}}
\def\bqu{\begin{question}}
	\def\equ{\end{question}}
\def\br{\begin{remark}}
	\def\er{\end{remark}}
\def\bt{\begin{theorem}}
	\def\et{\end{theorem}}

\def\squareforqed{\hbox{\rlap{$\sqcap$}$\sqcup$}}
\def\qed{\ifmmode\squareforqed\else{\unskip\nobreak\hfil
		\penalty50\hskip1em\null\nobreak\hfil\squareforqed
		\parfillskip=0pt\finalhyphendemerits=0\endgraf}\fi}
\def\endenv{\ifmmode\;\else{\unskip\nobreak\hfil
		\penalty50\hskip1em\null\nobreak\hfil\;
		\parfillskip=0pt\finalhyphendemerits=0\endgraf}\fi}
\newenvironment{proof}{\noindent \textbf{{Proof.~} }}{\qed}
\def\Dbar{\leavevmode\lower.6ex\hbox to 0pt
	{\hskip-.23ex\accent"16\hss}D}
\makeatletter
\def\url@leostyle{%
	\@ifundefined{selectfont}{\def\UrlFont{\sf}}{\def\UrlFont{\small\ttfamily}}}
\makeatother
\def\bcj{\begin{conjecture}}
	\def\ecj{\end{conjecture}}
\def\bcr{\begin{corollary}}
	\def\ecr{\end{corollary}}
\def\bd{\begin{definition}}
	\def\ed{\end{definition}}
\def\bea{\begin{eqnarray}}
	\def\eea{\end{eqnarray}}
\def\bem{\begin{enumerate}}
	\def\eem{\end{enumerate}}
\def\bex{\begin{example}}
	\def\eex{\end{example}}
\def\bim{\begin{itemize}}
	\def\eim{\end{itemize}}
\def\bl{\begin{lemma}}
	\def\el{\end{lemma}}
\def\bpf{\begin{proof}}
	\def\epf{\end{proof}}
\def\bpp{\begin{proposition}}
	\def\epp{\end{proposition}}
\def\bqu{\begin{question}}
	\def\equ{\end{question}}
\def\br{\begin{remark}}
	\def\er{\end{remark}}
\def\bt{\begin{theorem}}
	\def\et{\end{theorem}}

\def\btb{\begin{tabular}}
	\def\etb{\end{tabular}}

	\newcommand{\nc}{\newcommand}
	
	\nc{\bbA}{\mathbb{A}} \nc{\bbB}{\mathbb{B}} \nc{\bbC}{\mathbb{C}}
	\nc{\bbD}{\mathbb{D}} \nc{\bbE}{\mathbb{E}} \nc{\bbF}{\mathbb{F}}
	\nc{\bbG}{\mathbb{G}} \nc{\bbH}{\mathbb{H}} \nc{\bbI}{\mathbb{I}}
	\nc{\bbJ}{\mathbb{J}} \nc{\bbK}{\mathbb{K}} \nc{\bbL}{\mathbb{L}}
	\nc{\bbM}{\mathbb{M}} \nc{\bbN}{\mathbb{N}} \nc{\bbO}{\mathbb{O}}
	\nc{\bbP}{\mathbb{P}} \nc{\bbQ}{\mathbb{Q}} \nc{\bbR}{\mathbb{R}}
	\nc{\bbS}{\mathbb{S}} \nc{\bbT}{\mathbb{T}} \nc{\bbU}{\mathbb{U}}
	\nc{\bbV}{\mathbb{V}} \nc{\bbW}{\mathbb{W}} \nc{\bbX}{\mathbb{X}}
	\nc{\bbZ}{\mathbb{Z}}
	
	\nc{\bA}{{\bf A}} \nc{\bB}{{\bf B}} \nc{\bC}{{\bf C}}
	\nc{\bD}{{\bf D}} \nc{\bE}{{\bf E}} \nc{\bF}{{\bf F}}
	\nc{\bG}{{\bf G}} \nc{\bH}{{\bf H}} \nc{\bI}{{\bf I}}
	\nc{\bJ}{{\bf J}} \nc{\bK}{{\bf K}} \nc{\bL}{{\bf L}}
	\nc{\bM}{{\bf M}} \nc{\bN}{{\bf N}} \nc{\bO}{{\bf O}}
	\nc{\bP}{{\bf P}} \nc{\bQ}{{\bf Q}} \nc{\bR}{{\bf R}}
	\nc{\bS}{{\bf S}} \nc{\bT}{{\bf T}} \nc{\bU}{{\bf U}}
	\nc{\bV}{{\bf V}} \nc{\bW}{{\bf W}} \nc{\bX}{{\bf X}}
	\nc{\ba}{{\bf a}} \nc{\be}{{\bf e}} \nc{\bu}{{\bf u}}
	\nc{\brr}{{\bf r}} \nc{\bx}{{\bf x}} \nc{\bi}{{\bf i}}
	
	\nc{\cA}{{\cal A}} \nc{\cB}{{\cal B}} \nc{\cC}{{\cal C}}
	\nc{\cD}{{\cal D}} \nc{\cE}{{\cal E}} \nc{\cF}{{\cal F}}
	\nc{\cG}{{\cal G}} \nc{\cH}{{\cal H}} \nc{\cI}{{\cal I}}
	\nc{\cJ}{{\cal J}} \nc{\cK}{{\cal K}} \nc{\cL}{{\cal L}}
	\nc{\cM}{{\cal M}} \nc{\cN}{{\cal N}} \nc{\cO}{{\cal O}}
	\nc{\cP}{{\cal P}} \nc{\cQ}{{\cal Q}} \nc{\cR}{{\cal R}}
	\nc{\cS}{{\cal S}} \nc{\cT}{{\cal T}} \nc{\cU}{{\cal U}}
	\nc{\cV}{{\cal V}} \nc{\cW}{{\cal W}} \nc{\cX}{{\cal X}}
	\nc{\cZ}{{\cal Z}}
	
	\nc{\hA}{{\hat{A}}} \nc{\hB}{{\hat{B}}} \nc{\hC}{{\hat{C}}}
	\nc{\hD}{{\hat{D}}} \nc{\hE}{{\hat{E}}} \nc{\hF}{{\hat{F}}}
	\nc{\hG}{{\hat{G}}} \nc{\hH}{{\hat{H}}} \nc{\hI}{{\hat{I}}}
	\nc{\hJ}{{\hat{J}}} \nc{\hK}{{\hat{K}}} \nc{\hL}{{\hat{L}}}
	\nc{\hM}{{\hat{M}}} \nc{\hN}{{\hat{N}}} \nc{\hO}{{\hat{O}}}
	\nc{\hP}{{\hat{P}}} \nc{\hR}{{\hat{R}}} \nc{\hS}{{\hat{S}}}
	\nc{\hT}{{\hat{T}}} \nc{\hU}{{\hat{U}}} \nc{\hV}{{\hat{V}}}
	\nc{\hW}{{\hat{W}}} \nc{\hX}{{\hat{X}}} \nc{\hZ}{{\hat{Z}}}
	
	\nc{\hn}{{\hat{n}}}
	
    \def\ca{\mathop{\rm CA}}
    \def\can{\mathop{\rm CAN}}

	\def\max{\mathop{\rm max}}
	\def\min{\mathop{\rm min}}

	\def\tr{\mathop{\rm Tr}}

	\usepackage[
	colorlinks,
	linkcolor = blue,
	citecolor = blue,
	urlcolor = blue]{hyperref}
	\def \qed {\hfill \vrule height7pt width 7pt depth 0pt}
	
	\newcounter{lastnote}

\begin{document}
		\title{Optimal qudit overlapping tomography and optimal measurement order}
\author{Shuowei Ma}
	\affiliation{School of Computer Science and Engineering, Sun Yat-sen University, Guangzhou 510006, China}	

\author{Qianfan Wang}
	\affiliation{Department of Computer Science, City University of Hong Kong, Hong Kong SAR, China}

\author{Lvzhou Li}
	\affiliation{School of Computer Science and Engineering, Sun Yat-sen University, Guangzhou 510006, China}

\author{Fei Shi}
   \email[]{shif26@mail.sysu.edu.cn}
	\affiliation{School of Computer Science and Engineering, Sun Yat-sen University, Guangzhou 510006, China}

\begin{abstract}
Quantum state tomography is essential for characterizing quantum systems, but it becomes infeasible for large systems due to exponential resource scaling. Overlapping tomography addresses this challenge by reconstructing all $k$-body marginals using few measurement settings, enabling the efficient extraction of key information for many quantum tasks. While optimal schemes are known for qubits, the extension to higher-dimensional qudit systems remains largely unexplored. Here, we investigate optimal qudit overlapping tomography, constructing local measurement settings from generalized Gell-Mann matrices. By establishing a correspondence with combinatorial covering arrays, we present two explicit constructions of optimal measurement schemes. For $n$-qutrit systems, we prove that pairwise tomography requires at most $8 + 56\left\lceil \log_{8} n \right\rceil$ measurement settings, and provide an explicit scheme achieving this bound. Furthermore, we develop an efficient algorithm to determine the optimal order of these measurement settings, minimizing the experimental overhead associated with switching configurations. Compared to the worst-case ordering, our optimized schedule reduces switching costs by approximately 50\%. These results provide a practical pathway for efficient characterization of qudit systems, facilitating their application in quantum communication and computation.

  \end{abstract}
	\maketitle

\section{Introduction}\label{sec:int}

Quantum state tomography (QST) is a fundamental tool in quantum information science, enabling the reconstruction of unknown quantum states through a series of measurements \cite{PhysRevA.40.2847,RevModPhys.81.299,d2003quantum}. However, traditional QST becomes prohibitively resource-intensive as the system size increases, requiring an exponential number of measurements to fully characterize the state. In contrast, many quantum information tasks require only information about marginal states rather than the full quantum state \cite{PhysRevLett.89.207901,PhysRevA.69.022308,PhysRevX.8.021026,Eisert_2015,Cramer_2010}.  Overlapping tomography focuses on reconstructing all $k$-body marginals using a small number of local measurement settings \cite{PhysRevLett.124.100401}. This approach is particularly useful for measuring many-body entanglement and correlation functions \cite{BERRADA2022112621}, diagnosing long-range order and critical behavior, characterizing topological order \cite{PhysRevLett.132.240802,PhysRevLett.132.240801,PhysRevA.106.062441}, and estimating expectation values of $k$-local Hamiltonians \cite{Wang_2017,evans2019scalablebayesianhamiltonianlearning,Gebhart_2023,PhysRevLett.130.200403}.

Optimal qubit overlapping tomography seeks to determine the minimum number of Pauli measurement settings required to reconstruct all $k$-body marginals \cite{colbourn2010crc,t6qb-kdcp,PhysRevA.106.062441,PhysRevLett.130.050804,z2d4-3r8z}. This problem is closely connected to the covering number of covering arrays \cite{t6qb-kdcp}—a combinatorial structure widely used in experimental design and testing \cite{colbourn2004combinatorial,hedayat2012orthogonal,1321063,kuhn2013introduction}.
While most research on optimal overlapping tomography has focused on qubit systems, optimal qudit overlapping tomography remains largely unexplored. Qudit systems, which generalize qubits to $d$-level quantum systems, support more complex protocols and can significantly enhance the security and capacity of quantum communication \cite{PhysRevLett.88.127901,PhysRevLett.88.127902,PhysRevLett.85.3313,Etcheverry_2013,Erhard_2020}. Efficient tomography is essential for harnessing these advantages in practical applications.

Additionally, switching between local measurement settings in  experiments incurs significant overhead, as reconfiguring optical elements, laser polarizations, or microwave pulses often takes microseconds to milliseconds \cite{PhysRevA.93.012109,PhysRevA.96.032307,PRXQuantum.3.040310,PhysRevA.103.012420,PhysRevX.8.021012}. Moreover, frequent switching can increase control noise, crosstalk, and decoherence. Therefore, optimizing the measurement order of these local measurement settings is crucial.

In this work, we investigate optimal qudit overlapping tomography, in which local measurement settings are constructed from generalized Gell-Mann (GGM) matrices. Motivated by the correspondence between optimal qubit tomography and covering arrays \cite{t6qb-kdcp}, we employ covering arrays to present two explicit constructions of optimal qudit overlapping tomography schemes. For  $n$-qutrit systems, we prove that pairwise tomography requires at most $8 + 56\left\lceil \log_{8} n \right\rceil$ GGM measurement settings, and we provide an explicit scheme that achieves this bound. Furthermore, we present an efficient algorithm to determine the optimal measurement order. Compared to the worst-case measurement order, our optimal measurement scheme can reduce switching costs by approximately 50\%.

The rest of this paper is organized as follows. In Sec.~\ref{sec:pri}
, we introduce optimal qudit overlapping tomography, covering arrays, and the relationship between them. Sec.~\ref{sec:construction}
 presents two explicit constructions of optimal qudit overlapping tomography. In Sec.~\ref{sec:upper_bound}, we show several upper bounds for optimal qudit overlapping tomography. Sec.~\ref{sec:order}
 investigates the optimal measurement order. Finally, conclusions are provided in Sec.~\ref{sec:con}.

\section{Optimal qudit overlapping tomography and covering array}\label{sec:pri}



In this section, we discuss optimal qudit overlapping tomography, covering arrays, and the relationship between them. For qudit tomography, local measurements are typically performed using the 
GGM matrices \cite{PhysRevA.66.012303}, which is defined as follows \cite{Kimura_2003,Bertlmann_2008}:
\begin{itemize}
    \item[(i)] $\frac{d(d-1)}{2}$ symmetric GGM matrices:
    \[
    \Lambda_s^{jk} = |j\rangle\langle k| + |k\rangle\langle j|, \quad 1 \leq j < k \leq d;
    \]
    
    \item[(ii)] $\frac{d(d-1)}{2}$ antisymmetric GGM matrices:
    \[
    \Lambda_a^{jk} = -\mathrm{i}|j\rangle\langle k| + \mathrm{i}|k\rangle\langle j|, \quad 1 \leq j < k \leq d;
    \]
    
    \item[(iii)] $(d - 1)$ diagonal GGM matrices:
    \[\begin{aligned}
   \Lambda^l  =& \sqrt{\frac{2}{l(l+1)}} \left( \sum_{j=1}^l |j\rangle\langle j| - |l+1\rangle\langle l+1| \right),\\ 
    &1 \leq l \leq d - 1.
    \end{aligned}
    \]
\end{itemize}
When $d=2$, the GGM matrices correspond to the Pauli matrices. For simplicity, we denote the $d^2-1$
 GGM matrices as $\{\lambda_i\}_{i=1}^{d^2-1}$
 and use $\lambda_0$
 to represent the identity matrix of order $d$. Then a $d$-dimensional state $\varrho$ can be written as:
 \begin{equation}
    \varrho=\sum_{i\in \bbZ_{d^2}}a_i\lambda_i, 
 \end{equation}
where $a_0=\frac{1}{d}$, and $a_{i}=\frac{1}{2}\tr(\lambda_i\varrho)$. Similarly,  an $n$-qudit state $\varrho$ can be written as 
 \begin{equation} \varrho=\sum_{(i_1,i_2,\ldots,i_n)\in\bbZ_{d^2}^n}a_{i_1,i_2,\ldots,i_n}\lambda_{i_1}\lambda_{i_2} \cdots \lambda_{i_n},
 \end{equation}
where the tensor product symbols between the matrices $\lambda_{i_k}$ $(k=1,2,\ldots,n)$ are omitted, and $a_{i_1,i_2,\ldots,i_n}=\frac{1}{d^{n-t}2^t}\tr(\lambda_{i_1}\lambda_{i_2} \cdots \lambda_{i_n}\varrho)$
if there are $t$ nonzero elements in $(i_1,i_2,\ldots,i_n)$. Thus a full tomography of the $n$-qudit state $\varrho$ can be performed with $(d^2-1)^n$ GGM measurement settings $\{\lambda_{i_1}\lambda_{i_2} \cdots \lambda_{i_n}\mid i_1i_2\cdots i_n\neq 0, \,(i_1,i_2,\ldots,i_n)\in\bbZ_{d^2}^n\}$.

The goal of qudit overlapping tomography is to reconstruct all $k$-body marginals of an arbitrary $n$-qudit state. Obviously, conventional quantum state tomography would require $(d^2-1)^k \binom{n}{k}$ distinct GGM measurement settings~\cite{PhysRevA.96.032307,PhysRevA.66.012303,PhysRevA.64.052312}. However, this method involves substantial redundancy, since many measurement settings are effectively repeated across different $k$-body subsystems. Thus it is important to construct optimal measurement schemes. Next, we introduce optimal qudit overlapping tomography.


\begin{definition}
   Optimal qudit overlapping tomography is the problem of determining the minimum number of GGM measurement settings, denoted by $\phi_k(n,d)$, required to cover all $k$-body marginals of an $n$-qudit state. 
\end{definition}

For example, in the case of four-qubit state, only nine Pauli measurement settings, namely XXXX, ZYYX, YZZX, YYXY, XZYY, ZXZY, ZZXZ, YXYZ, and XYZZ, are sufficient to reconstruct all 2-body marginals, i.e., $\phi_2(4,2)=9$ \cite{t6qb-kdcp}. This is because, for every pair of qubits, these nine combinations of Pauli operators collectively provide the necessary information for state reconstruction.
By  applying the  bijection: $X\rightarrow 0$, $Y\rightarrow 1$, $Z\rightarrow 2$, the nine Pauli measurement settings can be represented as a $9\times 4$ array,
\begin{equation}\label{eq:9times4}
 \begin{matrix}
0  &0   &0   &0\\
2  &1   &1   &0\\
1  &2   &2   &0\\
1  &1   &0   &1\\
0  &2   &1   &1\\
2  &0   &2   &1\\
2  &2   &0   &2\\
1  &0   &1   &2\\
0  &1   &2   &2\\  
\end{matrix},   
\end{equation}
 which has the property that in every $9\times 2$
 subarray, each possible $2$-tuple from $\bbZ_3\times \bbZ_3$
 appears exactly once. This property gives rise to a combinatorial object known as a covering array  \cite{colbourn2010crc}.

\begin{definition}
 A \emph{covering array} $\ca(r; k, n, v)$ is an $r \times n$ array with entries from an alphabet of size $v$, such that in every $r\times k$ subarrary, each possible $k$-tuple from $v\times v\times \cdots \times v$ appears  at least once. Given $k$, $n$, and $v$, the covering number $\can(k,n,v)$ is the minimum number of rows $r$ for which a $\ca(r; k, n, v)$ exists.
\end{definition}

The optimal qubit overlapping tomography using Pauli measurements corresponds to the covering number of covering arrays \cite{t6qb-kdcp}. Similarly, the optimal qudit overlapping tomography using GGM measurement settings also corresponds to the covering number of covering arrays.

\begin{observation}\label{observation}
The minimum number of GGM measurement settings required to reconstruct all 
$k$-body marginals of an 
$n$-qudit state is equal to the  minimum number of rows $r$
 for which a covering array $\ca(r; k, n, d^2-1)$
 exists, i.e,
    \begin{equation}
        \phi_{k}(n,d)=\can(k,n,d^2-1).
    \end{equation}
\end{observation}

Obviously,  $\phi_{k}(n,d)=\can(k,n,d^2-1)\geq (d^2-1)^k$.
Some tables of covering numbers can be found in Refs.~\cite{colbourn2010crc,catable2019}. However, only a small number of covering numbers have been determined explicitly, as the problem of finding the general covering number is NP-hard. In the next section, we present several representative constructions.

\section{Constructions of optimal qudit overlapping tomography}\label{sec:construction}

In this section, we present two explicit constructions of optimal qudit overlapping tomography based on covering arrays. The first construction is based on Zero-Sum construction \cite{colbourn2010crc}.

\begin{theorem}\label{pro:two}
\(  \phi_{k}(k+1,d) = (d^2 - 1)^k \).
\end{theorem}

The construction of a covering array  $\ca((d^2-1)^k; k, k+1, d^2-1)$ is as follows.
List all possible $k$-tuples over $\mathbb{Z}_{d^2-1}$. For each $k$-tuple $(a_1, a_2, \ldots, a_k)$, define the $(k+1)$-th element as $-(a_1 + a_2 + \cdots + a_k)$ (computed modulo $d^2-1$). In this way, we obtain a covering array $\ca((d^2-1)^k; k, k+1, d^2-1)$ consisting of the rows: 
\begin{equation}
 (a_1, a_2, \ldots, a_k, -(a_1 + a_2 + \cdots + a_k)).   
\end{equation}


    
    
    
For example, when $d=2$, $k=2$,  we obtain a covering array $\ca(9;2,3,3)$ by the above construction:
\begin{equation}\label{eq:array93}
 \begin{matrix}
0  &0   &0\\
0  &1   &2\\
0  &2   &1\\
1  &0   &2\\
1  &1   &1\\
1  &2   &0\\
2  &0   &1\\
2  &1   &0\\
2  &2   &2\\  
\end{matrix}.  
\end{equation}



The second construction is based on Bush's construction \cite{hedayat2012orthogonal}.

\begin{theorem}
\label{pro:three}
When $d^2 - 1 > k$, and $d^2 - 1$ is a prime power, then $ \phi_{k}(d^2,d) = (d^2 - 1)^k$.
\end{theorem}

Note that when $d^2-1$ is a prime power, the possible values for $d$
 are only $2$
 or $3$. Therefore, Theorem~\ref{pro:three}
 holds only for $2$
 or $3$.  Here, we need to use Galois field \cite{lidl1997finite}. A Galois field is a field consisting of a finite set of elements, in which addition, subtraction, multiplication, and division (except by zero) are defined and satisfy the field axioms. We usually denote 
$\text{GF}(s)$ as the Galois field with $s$
 elements, where $s$
 must be a prime power.


The construction of a covering array $\ca((d^2-1)^k;k,d^2,d^2-1)$
 is as follows. Consider all polynomials over the finite field $\text{GF}(d^2-1)$
 of degree at most $k-1$, that is, 
    $a_0+a_1x+a_2x^2+\cdots+a_{k-1}x^{k-1}$, where $a_i\in \text{GF}(d^2-1)$ for $0\leq i\leq k-1$. There are exactly $(d^2-1)^k$
 such polynomials, which we denote by $\{f_1, f_2, \ldots, f_{(d^2-1)^k}\}$. Then the covering array $\ca((d^2-1)^k;k,d^2,d^2-1)$ is constructed as follows: 

\begin{equation}
 \begin{matrix}
f_1(\alpha_1)  &f_1(\alpha_2)   &\cdots &f_1(\alpha_{d^2-1}) & r_1 \\
f_2(\alpha_1)  &f_2(\alpha_2)   &\cdots &f_2(\alpha_{d^2-1}) & r_2 \\
\vdots  &\vdots   &\vdots &\vdots &\vdots \\
f_{(d^2-1)^k}(\alpha_1)  &f_{(d^2-1)^k}(\alpha_2)   &\cdots &f_{(d^2-1)^k}(\alpha_{d^2-1}) & r_{(d^2-1)^k} \\
\end{matrix},  
\end{equation}
where $\alpha_1,\alpha_2,\ldots, \alpha_{d^2-1}$ are the $d^2-1$ elements of $\text{GF}(d^2-1)$, and $r_i$ is the coefficient of the 
term $x^{k-1}$ in the polynomial $f_i$ for $1\leq i\leq (d^2-1)^k$.

Next, we show a simple example. Let $d=2$ and $k=3$, then the Galois field $\text{GF}(3)=\{0, 1, 2\}$. The $9$ polynomials are $f_1=0$, $f_2=1$, $f_3=2$,  $f_4=x$, $f_5=1+x$, $f_6=2+x$, $f_7=2x$, $f_8=1+2x$, and $f_9=2+2x$. Then we obtain the covering array $\ca(9;2,4,3)$:
\begin{equation}
 \begin{matrix}
0  &0   &0   &0\\
1  &1   &1   &0\\
2  &2   &2   &0\\
0  &1   &2   &1\\
1  &2   &0   &1\\
2  &0   &1   &1\\
0  &2   &1   &2\\
1  &0   &2   &2\\
2  &1   &0   &2\\  
\end{matrix}.   
\end{equation}
When we perform row and column permutations on this array, we would obtain Eq.~\eqref{eq:9times4}.




\section{upper bounds on optimal qudit overlapping tomography }\label{sec:upper_bound}

In this section, we show some upper bounds on optimal qudit overlapping tomography.  For small $d$, $k$, and $n$, we obtain the best known upper bound on $\phi_k(n,d)$ from the handbook~\cite{colbourn2010crc} and the website~\cite{catable2019}, and see Table~\ref{tab:phi_values_d2}.




For general values of $k$, $n$, and $d$, there is no universal construction for the covering array $\ca(r; k, n, d^2-1)$ when $r$ is set to the best known upper bound for $\phi_k(n, d)$.
However, some upper bounds for $\phi_k(n, d)$ have been established using probabilistic methods.

\begin{enumerate}[(1)]
\item When $n \geq k \geq 2$ and $d \geq 2$, the following efficient upper bound holds: \cite{doi:10.1137/16M1067767}:
 \begin{equation}
     \begin{aligned}
   \phi_{k}(n,d) \leq &\frac{1}{\log \left( \frac{(d^2-1)^k}{(d^2-1)^{k}-1} \right)} \times \left[ \log \binom{n}{k} + k \log (d^2-1)\right. \\
  &\left.+ \log \log \left( \frac{(d^2-1)^k}{(d^2-1)^{k}-1} \right) + 1 \right].        
     \end{aligned}
 \end{equation}

 \item When \(n \to \infty\), \( n \geq k \geq 2 \), and \( d \geq 2 \), the following upper bound, known as the Stein-Lov\'asz-Johnson (SLJ) bound  holds ~\cite{10.1145/800125.804034,LOVASZ1975383,STEIN1974391}:
\begin{equation}
\phi_{k}(n,d) \leq \frac{k}{\log \frac{(d^2-1)^k}{(d^2-1)^k-1}} \log n \bigl(1+o(1)\bigr).
\end{equation}
\end{enumerate}



\begin{table}[t]
\renewcommand\arraystretch{0.8}	
	\centering
	\renewcommand\tabcolsep{3pt}
\caption{The best known upper bounds on $\phi_k(n,2)$ for $2\leq k\leq 6$ and $4\leq n\leq 20$, and on $\phi_k(n,3)$ for $2\leq k\leq 6$ and $8\leq n\leq 20$.}
\label{tab:phi_values_d2}
\begin{tabular}{c|ccccc}
\toprule
$n$ & $\phi_2(n,2)$ & $\phi_3(n,2)$ & $\phi_4(n,2)$ & $\phi_5(n,2)$ & $\phi_6(n,2)$ \\
\midrule
4   &9                &27                &81                &\textemdash                &\textemdash                \\
5   &11                &33                &81                &243                &\textemdash                \\
6   &12                &33                &111                &243                &729                \\
7   &12                &39                &123                &351                &729                \\
8   &13                &42                &135                &405                &1134                \\
9   &13                &45                &135                &405                &1377                \\
10  &14                &45                &159                &405                &1431                \\
11  &15                &45                &159                &483                &1431                \\
12  &15                &45                &189                &483                &1455                \\
13  &15                &45                &212                &687                &2181                \\
14  &15                &45                &231                &805                &2701                \\
15  &15                &51                &231                &842                &2901                \\
16  &15                &51                &237                &920                &3126                \\
17  &15                &58                &237                &963                &3633                \\
18  &15                &59                &271                &1034                &3839                \\
19  &15                &59                &271                &1064                &3961                \\
20  &15                &59                &271                &1108                &4006                \\
\bottomrule 

$n$ & $\phi_2(n,3)$ & $\phi_3(n,3)$ & $\phi_4(n,3)$ & $\phi_5(n,3)$ & $\phi_6(n,3)$ \\
\midrule
8  &64                &512                &4096                &32768                &262144                \\
9   &64                &512                &4096                &32768                &262144                \\
10   &76                &512                &6125                &53681                &450372                \\
11   &78                &960                &7680               &61440                &450372                \\
12   &84                &960                 &7680                &61440                &491520                \\
13   &84                &960                 &7680                &61440                &520192                \\
14   &96                &960                 &7680                &65024                &753656               \\
15  &96                &960                 &7680                &65024                &753656                \\
16  &102                &960                &8128                &94200                &753656                \\
17  &104                &960                &8128                &94200                &753656                \\
18  &104                &960                 &8128                &94200                &782328                \\
19  &107                &1016                &8128               &94200                &983032               \\
20  &108                &1016                &8184                &94200                &983032              \\
\bottomrule
\end{tabular}
\end{table}

Pairwise tomography ($k=2$) is the most common form of overlapping tomography.
Ref.~\cite{PhysRevResearch.2.023393} provides an efficient upper bound for pairwise qubit tomography, $\phi_2(n,2) \leq 3 + 6\left\lceil\log_{3} n \right\rceil$, along with an explicit construction.  Next, we generalize this result to the qutrit system.

\begin{theorem}\label{pro:qutrit_overlaping}
    In an $n$-qutrit system, 
    \begin{equation}
       \phi_2(n,3)\leq 8 + 56\left\lceil \log_{8} n \right\rceil. 
    \end{equation}
\end{theorem}





Theorem~\ref{pro:qutrit_overlaping} shows that, for $n$-qutrit overlapping tomography, the number of GGM measurement settings grows only logarithmically with respect to $n$. To explicitly construct these GGM measurement settings, it suffices to construct a covering array $\ca(8+ 56\left\lceil \log_{8} n \right\rceil; 2, n, 8)$. Note that, we use $\bi_n$ to denote a row vector of length $n$, in which every element is $i$.  There are three steps:

1. We need to use $\ca(64; 2, 8, 8)$ (see Table~\ref{oa:8oa} in Appendix~\ref{app:arr}) which contains rows $\{r_i=\bi_8\}_{i=0}^7$, and the remaining rows are denoted by $\{r_i\}_{i=8}^{63}$. 

2. Next, we need to represent $0, 1, 2, \ldots, n-1$ in base-$8$ notation. Each element can be written as a column vector of length $\lceil \log_{8} n \rceil$. In this way, we obtain an array $A$ of size $\lceil \log_{8} n \rceil \times n$. For example, when $n = 10$, we obtain a $2 \times 10$  array:
\begin{equation}
 A=\begin{matrix}
0   &0   &0   &0   &0   &0   &0   &0   &1   &1\\
0   &1   &2   &3   &4   &5   &6   &7   &0   &1\\
\end{matrix}.   
\end{equation}

3.  For each $r_{i}$ ($8\leq i\leq 63$),   we replace every element in $A$ as follows: $j\rightarrow r_{i,j}$,
where $0\leq j\leq 7$, and $r_{i,j}$ is the element in the $j$-th column of $r_i$. Then we obtain a new $\lceil \log_{8} n \rceil \times n$ array $A_{r_i}$. Now the covering array $\ca(8+ 56\left\lceil \log_{8} n \right\rceil; 2, n, 8)$ can be constructed as follows:
\begin{equation}
\begin{matrix}
     \textbf{0}_n\\
    \textbf{1}_n\\
    \vdots\\
      \textbf{7}_n\\
     A_{r_8}\\
        A_{r_{9}}\\
    \vdots\\
      A_{r_{63}}\\
\end{matrix}.
\end{equation}

According to the above construction, we have the following corollary. 
\begin{corollary}
    If there exists a covering array $\ca((d^2-1)^2;2,d^2-1,d^2-1)$ that contains the rows $\{\textbf{i}_{d^2-1}\}_{i=0}^{d^2-2}$, then 
    \begin{equation}
       \phi_2(n,d)\leq (d^2 - 1) + (d^2 - 1)(d^2-2)\left\lceil \log_{d^2 - 1} n \right\rceil. 
    \end{equation}
\end{corollary}

Note that when $d = 2$, there exists a covering array $\ca(9; 2, 3, 3)$ that contains the rows $\{\textbf{i}_3\}_{i=0}^{2}$ (see Eq.~\eqref{eq:array93}
). Therefore, $\phi_2(n, 2) \leq 3 + 6\left\lceil \log_{3} n \right\rceil$ can be obtained in this way~\cite{PhysRevResearch.2.023393}.

\section{Optimal measurement order}\label{sec:order}



In this section, we consider the problem of determining the optimal measurement order, which minimizes the total overhead associated with switching between measurement settings.

When performing a sequence of local measurements in quantum experiments, switching between different measurement settings is not instantaneous and introduces a non-negligible overhead. Adjustments such as rotating optical elements, tuning laser polarization, or reconfiguring microwave pulses typically require timescales ranging from microseconds to milliseconds \cite{PhysRevA.93.012109,PhysRevA.96.032307,PRXQuantum.3.040310,PhysRevA.103.012420,PhysRevX.8.021012}. Furthermore, frequent switching increases the risk of control noise, crosstalk, and decoherence. Therefore, given a set of required Pauli or GGM basis measurement settings, determining the order of execution that minimizes the total switching cost becomes a sequence optimization problem.

The Hamming distance between two vectors is defined as the number of positions at which their corresponding entries differ \cite{6772729,macwilliams1977theory}. Here, we use the Hamming distance to quantify the switching cost between two local measurements.

\begin{table}[t]
  \centering
  \small
  \caption{Minimum cost and maximum cost for CA (33,3,6,3).}
  \label{tab:ms4_example}
  \setlength{\tabcolsep}{3pt}  
  \renewcommand{\arraystretch}{0.5}  
   \setlength{\extrarowheight}{-1pt}     
  \begin{tabular}{ccccccc}
    \toprule
    0 & 1 & 2 & 2 & 1 & 0 &  \\
    0 & 1 & 2 & 0 & 2 & 1 & (cost: 3) \\
    0 & 2 & 2 & 1 & 0 & 1 & (cost: 3) \\
    0 & 0 & 2 & 1 & 1 & 2 & (cost: 3) \\
    0 & 1 & 0 & 1 & 2 & 2 & (cost: 3) \\
    0 & 1 & 1 & 2 & 0 & 2 & (cost: 3) \\
    0 & 0 & 1 & 2 & 2 & 1 & (cost: 3) \\
    1 & 0 & 2 & 2 & 0 & 1 & (cost: 3) \\
    2 & 1 & 0 & 2 & 0 & 1 & (cost: 3) \\
    0 & 2 & 0 & 2 & 1 & 1 & (cost: 3) \\
    0 & 2 & 1 & 0 & 1 & 2 & (cost: 3) \\
    1 & 0 & 1 & 0 & 2 & 2 & (cost: 3) \\
    1 & 0 & 2 & 1 & 2 & 0 & (cost: 3) \\
    1 & 1 & 0 & 2 & 2 & 0 & (cost: 3) \\
    2 & 1 & 1 & 0 & 2 & 0 & (cost: 3) \\
    0 & 2 & 1 & 1 & 2 & 0 & (cost: 3) \\
    2 & 2 & 0 & 1 & 1 & 0 & (cost: 3) \\
    1 & 2 & 0 & 1 & 0 & 2 & (cost: 3) \\
    1 & 0 & 0 & 2 & 1 & 2 & (cost: 3) \\
    1 & 0 & 1 & 1 & 0 & 2 & (cost: 3) \\
    1 & 1 & 1 & 1 & 1 & 1 & (cost: 3) \\
    2 & 2 & 1 & 0 & 0 & 1 & (cost: 4) \\
    1 & 2 & 0 & 0 & 2 & 1 & (cost: 3) \\
    1 & 2 & 2 & 0 & 1 & 0 & (cost: 3) \\
    2 & 0 & 2 & 0 & 1 & 1 & (cost: 3) \\
    2 & 1 & 0 & 0 & 1 & 2 & (cost: 3) \\
    1 & 1 & 2 & 0 & 0 & 2 & (cost: 3) \\
    2 & 1 & 2 & 1 & 0 & 0 & (cost: 3) \\
    0 & 1 & 0 & 1 & 0 & 0 & (cost: 2) \\
    0 & 0 & 0 & 0 & 0 & 0 & (cost: 2) \\
    2 & 0 & 0 & 1 & 2 & 1 & (cost: 4) \\
    2 & 0 & 1 & 2 & 1 & 0 & (cost: 4) \\
    2 & 2 & 2 & 2 & 2 & 2 & (cost: 4) \\
    \midrule
    \multicolumn{6}{r}{Minimum cost:} & 98 \\
    \bottomrule
  \end{tabular}
  \quad \quad 
  \begin{tabular}{ccccccc}
    \toprule
    0 & 1 & 2 & 2 & 1 & 0 &  \\
    1 & 0 & 1 & 1 & 0 & 2 & (cost: 6) \\
    1 & 1 & 0 & 2 & 2 & 0 & (cost: 5) \\
    2 & 2 & 1 & 0 & 0 & 1 & (cost: 6) \\
    0 & 0 & 2 & 1 & 1 & 2 & (cost: 6) \\
    1 & 2 & 0 & 0 & 2 & 1 & (cost: 6) \\
    2 & 1 & 2 & 1 & 0 & 0 & (cost: 6) \\
    1 & 0 & 1 & 0 & 2 & 2 & (cost: 6) \\
    0 & 1 & 0 & 1 & 0 & 0 & (cost: 6) \\
    2 & 0 & 2 & 0 & 1 & 1 & (cost: 6) \\
    0 & 1 & 0 & 1 & 2 & 2 & (cost: 6) \\
    1 & 0 & 2 & 2 & 0 & 1 & (cost: 6) \\
    2 & 1 & 1 & 0 & 2 & 0 & (cost: 6) \\
    0 & 2 & 2 & 1 & 0 & 1 & (cost: 6) \\
    1 & 0 & 0 & 2 & 1 & 2 & (cost: 6) \\
    0 & 1 & 2 & 0 & 2 & 1 & (cost: 6) \\
    2 & 2 & 0 & 1 & 1 & 0 & (cost: 6) \\
    0 & 0 & 1 & 2 & 2 & 1 & (cost: 6) \\
    1 & 1 & 2 & 0 & 0 & 2 & (cost: 6) \\
    0 & 2 & 0 & 2 & 1 & 1 & (cost: 6) \\
    1 & 0 & 2 & 1 & 2 & 0 & (cost: 6) \\
    0 & 2 & 1 & 0 & 1 & 2 & (cost: 6) \\
    2 & 1 & 0 & 2 & 0 & 1 & (cost: 6) \\
    0 & 2 & 1 & 1 & 2 & 0 & (cost: 6) \\
    2 & 1 & 0 & 0 & 1 & 2 & (cost: 6) \\
    0 & 1 & 1 & 2 & 0 & 2 & (cost: 4) \\
    1 & 2 & 2 & 0 & 1 & 0 & (cost: 6) \\
    2 & 0 & 0 & 1 & 2 & 1 & (cost: 6) \\
    1 & 2 & 0 & 1 & 0 & 2 & (cost: 4) \\
    2 & 0 & 1 & 2 & 1 & 0 & (cost: 6) \\
    0 & 0 & 0 & 0 & 0 & 0 & (cost: 4) \\
    1 & 1 & 1 & 1 & 1 & 1 & (cost: 6) \\
    2 & 2 & 2 & 2 & 2 & 2 & (cost: 6) \\
    \midrule
    \multicolumn{6}{r}{Maximum cost:} & 185 \\
    \bottomrule
  \end{tabular}
  \setlength{\tabcolsep}{6pt}  
  \renewcommand{\arraystretch}{1.0}
\end{table}

\begin{definition}
Give two local GGM measurement settings $M_i=\lambda_{i_1}\lambda_{i_2}\cdots \lambda_{i_n}$ and $M_j=\lambda_{j_1}\lambda_{j_2}\cdots \lambda_{j_n}$, the
 switching cost between two measurement settings is defined as:
\begin{equation}
  d(M_i, M_j)=|\{k|\lambda_{i_k}\neq \lambda_{j_k}, \, \forall\, 1\leq k\leq n\}|. 
\end{equation}
\end{definition}

Given a set of GGM  measurements settings, $\{M_i\}_{i=1}^m$, our goal is to determine the optimal measurement order that minimizes the total switching cost between these measurements. Specifically, we need to find a permutation $\pi: [m] \to [m]$ that minimizes the total cost:
\begin{align}
\text{Total cost} = \sum_{i=1}^{m-1} d(M_{\pi(i)}, M_{\pi(i+1)}).
\end{align}
The optimization problem can be formulated as:
\begin{align}
\text{Minimum cost} =\min_{\pi \in S_m} \sum_{i=1}^{m-1} d(M_{\pi(i)}, M_{\pi(i+1)})
\end{align}
where \( S_m \) is the set of all permutations of \( m \) elements.

This problem can be essentially transformed into the shortest Hamiltonian path problem on a complete weighted graph \( K_m \), where each node \( v_i \) corresponds to the \( i \)-th measurement configuration \( M_i \), and the weight of edge \( e_{ij} \) is \( w_{ij} = d(M_i, M_j) \).

Various methods have been studied for the shortest Hamiltonian path problem \cite{doi:10.1137/0110015,10.1016/j.dam.2021.11.019,10.1145/2818310,madkour2017surveyshortestpathalgorithms}. While dynamic programming is often used, considering that the number of qudits and measurement settings might be very large, we propose a hybrid algorithm combining dynamic programming (for small-scale problems), an improved cluster-based nearest neighbor + 2-opt heuristic algorithm (for medium-scale problems), and simulated annealing (for large-scale problems). The specific algorithm or pseudocode is provided in Appendix~\ref{appendix:algorithm}. Below, we give an example.


We consider the covering array $\ca(33;3,6,3)$, which corresponds to the construction of the optimal qudit overlapping tomography with $\phi_3(6,2)=33$. Table~\ref{tab:ms4_example}
shows both the optimal measurement order along with its minimum cost, as well as the worst measurement order with its maximum cost. The algorithm used to determine the maximum cost is similar to the optimization algorithm, but with the objective reversed.
By comparing these two tables, we observe that optimization reduces the total cost from 185 to 98, achieving an improvement of nearly $50\%$.

\begin{figure}\label{fig:subfig_b}
    \centering
        \includegraphics[scale=0.2]{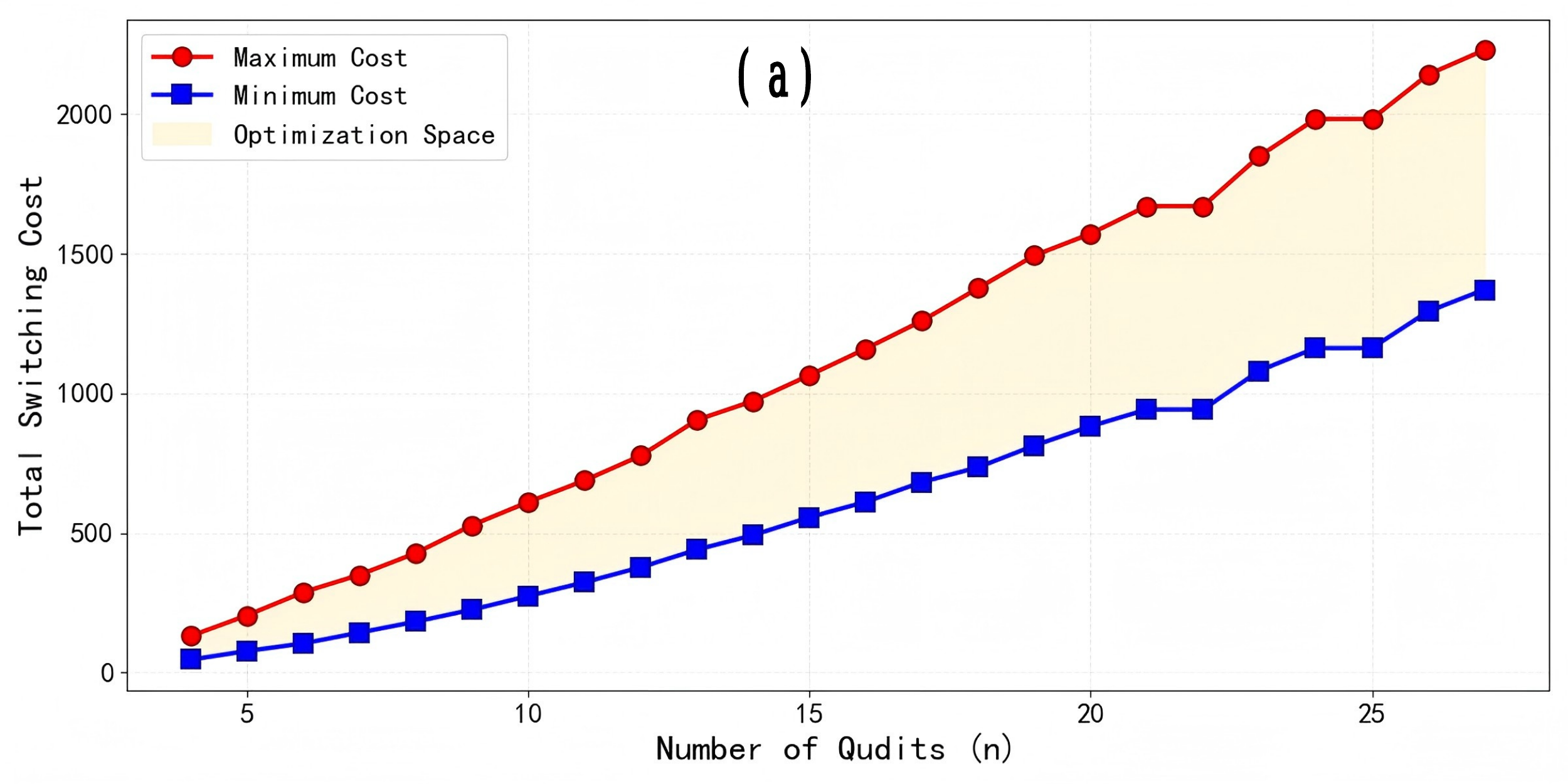}
        \includegraphics[scale=0.2]{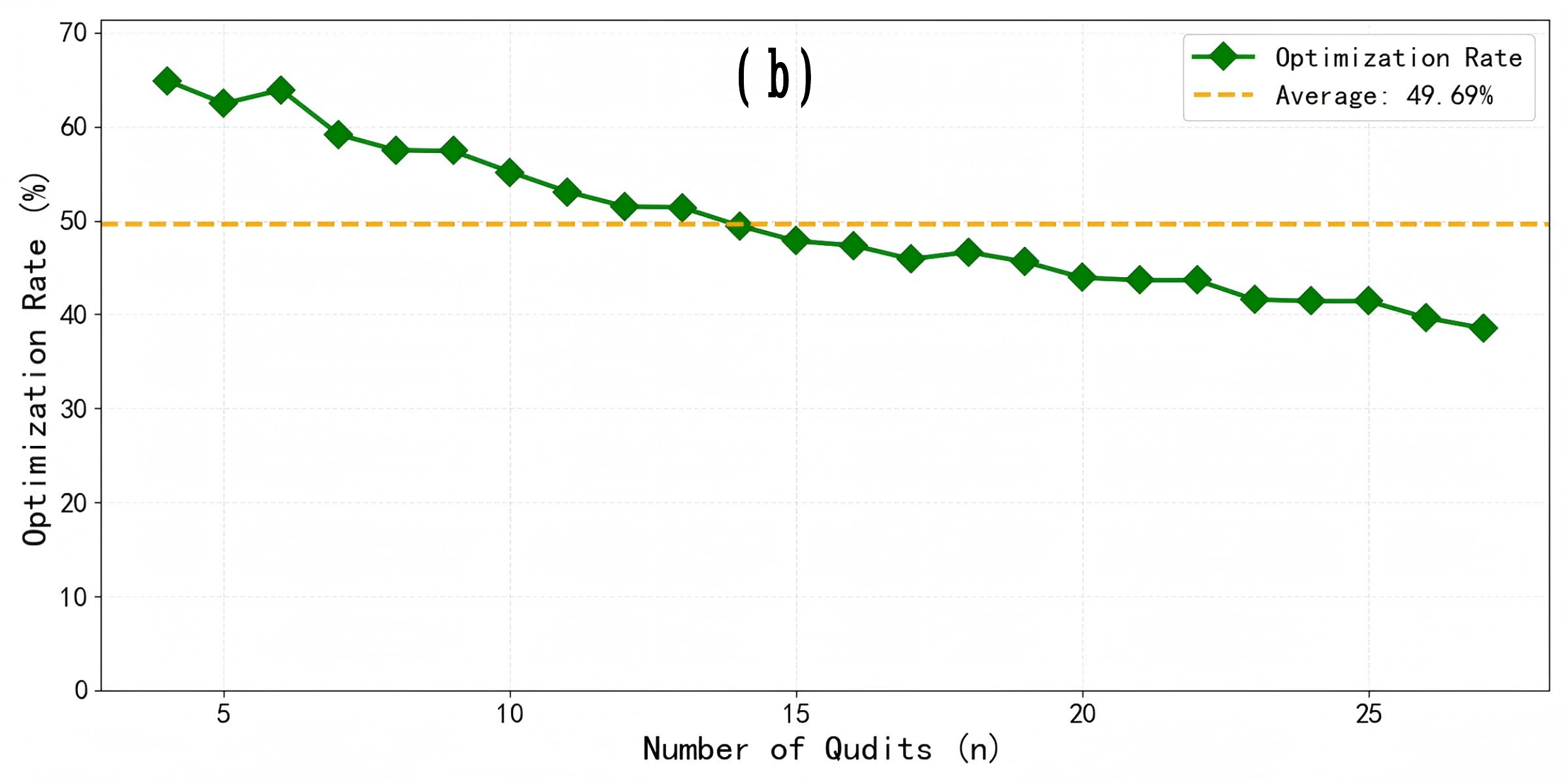}
    \caption{
        \textbf{Comparison between the optimal measurement order and the worst measurement order.} (a). The red line represents the minimum cost achieved by the optimal measurement order, while the blue line represents the maximum cost corresponding to the worst measurement order, as the number of qudits increases from $4$ to $27$. The yellow region indicates the portion of cost that is reduced through optimization. (b) The green line represents the optimization rate, which is calculated as $\frac{\mathrm{Maximum\ cost} - \mathrm{Minimum\ cost}}{\mathrm{Maximum\ cost}} \times 100\%$.}
    \label{fig:com_figure}
\end{figure}

Since there is currently no program or method capable of constructing all types of small-scale optimal covering arrays, we can instead use the IPOG-F algorithm from \cite{forbes2008refining} to generate near-optimal covering arrays for our experiments.
Based on the IPOG-F algorithm, we generated 24 covering arrays with fixed \(d=2\), \(k=3\), and \(n\) ranging from 4 to 27. Using these covering arrays, we conducted simulation experiments on a computer and plotted the maximum and minimum costs as functions of $n$ (ranging from 4 to 27), as shown in Figure~\ref{fig:com_figure}.  Compared to the worst-case ordering, our optimized schedule reduces switching costs by approximately 50\%, as indicated in the figure.


\section{Conclusions}\label{sec:con}
In this work, we have investigated optimal qudit overlapping tomography using GGM measurement settings. By employing covering arrays, we presented two constructions for optimal qudit overlapping tomography schemes. For 
$n$-qutrit systems, we established an upper bound of $8 + 56\left\lceil \log_{8} n \right\rceil$
 measurement settings for pairwise tomography and provided an explicit scheme that achieves this bound. Furthermore, we introduced an algorithm to optimize the measurement order, reducing experimental switching costs by approximately 50\% compared to the worst-case scenario.

Our results bridge a critical gap in quantum state characterization by enabling efficient overlapping tomography for qudit systems, which are increasingly relevant in quantum communication, error correction, and simulation. The reduction in measurement settings and optimized switching order not only to theoretical efficiency but also to practical feasibility in experimental implementations, where measurement reconfiguration overhead and noise are major concerns. Given that optimal pairwise overlapping tomography for six-qubit systems has already been implemented on optical platforms \cite{t6qb-kdcp}, realizing our proposed optimal overlapping qudit tomography scheme on experimental platforms is also of significant importance and research value. Another interesting direction is to study optimal overlapping tomography in mixed systems by using mixed covering arrays.


\section*{Acknowledgments}
\label{sec:ack}		
We thank Qi Zhao, Xingjian Zhang, Tianfeng Feng, Huan Cao, and  Dian Wu for discussing this problem. 

\bibliographystyle{apsrev4-2}
\clearpage
\bibliography{reference}

\appendix
\onecolumngrid

\section{$\ca(64;2,8,8)$}\label{app:arr}
\begin{table}[htbp]
\centering
\caption{$\ca(64;2,8,8)$}\label{oa:8oa}
\begin{tabular}{cccccccc}
$r_0$=$0$ & $0$ & $0$ & $0$ & $0$ & $0$ & $0$ & $0$ \\
$r_1$=$1$ & $1$ & $1$ & $1$ & $1$ & $1$ & $1$ & $1$ \\
$r_2$=$2$ & $2$ & $2$ & $2$ & $2$ & $2$ & $2$ & $2$ \\
$r_3$=$3$ & $3$ & $3$ & $3$ & $3$ & $3$ & $3$ & $3$ \\
$r_4$=$4$ & $4$ & $4$ & $4$ & $4$ & $4$ & $4$ & $4$ \\
$r_5$=$5$ & $5$ & $5$ & $5$ & $5$ & $5$ & $5$ & $5$ \\
$r_6$=$6$ & $6$ & $6$ & $6$ & $6$ & $6$ & $6$ & $6$ \\
$r_7$=$7$ & $7$ & $7$ & $7$ & $7$ & $7$ & $7$ & $7$ \\
$r_8$=$0$ & $1$ & $2$ & $3$ & $4$ & $5$ & $6$ & $7$ \\
$r_9$=$0$ & $2$ & $3$ & $4$ & $5$ & $6$ & $7$ & $1$ \\
$r_{10}$=$0$ & $3$ & $4$ & $5$ & $6$ & $7$ & $1$ & $2$ \\
$r_{11}$=$0$ & $4$ & $5$ & $6$ & $7$ & $1$ & $2$ & $3$ \\
$r_{12}$=$0$ & $5$ & $6$ & $7$ & $1$ & $2$ & $3$ & $4$ \\
$r_{13}$=$0$ & $6$ & $7$ & $1$ & $2$ & $3$ & $4$ & $5$ \\
$r_{14}$=$0$ & $7$ & $1$ & $2$ & $3$ & $4$ & $5$ & $6$ \\
$r_{15}$=$1$ & $0$ & $4$ & $7$ & $2$ & $6$ & $5$ & $3$ \\
\end{tabular}
\quad
\begin{tabular}{cccccccc}
$r_{16}$=$1$ & $4$ & $7$ & $2$ & $6$ & $5$ & $3$ & $0$ \\
$r_{17}$=$1$ & $7$ & $2$ & $6$ & $5$ & $3$ & $0$ & $4$ \\
$r_{18}$=$1$ & $2$ & $6$ & $5$ & $3$ & $0$ & $4$ & $7$ \\
$r_{19}$=$1$ & $6$ & $5$ & $3$ & $0$ & $4$ & $7$ & $2$ \\
$r_{20}$=$1$ & $5$ & $3$ & $0$ & $4$ & $7$ & $2$ & $6$ \\
$r_{21}$=$1$ & $3$ & $0$ & $4$ & $7$ & $2$ & $6$ & $5$ \\
$r_{22}$=$2$ & $4$ & $0$ & $5$ & $1$ & $3$ & $7$ & $6$ \\
$r_{23}$=$2$ & $0$ & $5$ & $1$ & $3$ & $7$ & $6$ & $4$ \\
$r_{24}$=$2$ & $5$ & $1$ & $3$ & $7$ & $6$ & $4$ & $0$ \\
$r_{25}$=$2$ & $1$ & $3$ & $7$ & $6$ & $4$ & $0$ & $5$ \\
$r_{26}$=$2$ & $3$ & $7$ & $6$ & $4$ & $0$ & $5$ & $1$ \\
$r_{27}$=$2$ & $7$ & $6$ & $4$ & $0$ & $5$ & $1$ & $3$ \\
$r_{28}$=$2$ & $6$ & $4$ & $0$ & $5$ & $1$ & $3$ & $7$ \\
$r_{29}$=$3$ & $7$ & $5$ & $0$ & $6$ & $2$ & $4$ & $1$ \\
$r_{30}$=$3$ & $5$ & $0$ & $6$ & $2$ & $4$ & $1$ & $7$ \\
$r_{31}$=$3$ & $0$ & $6$ & $2$ & $4$ & $1$ & $7$ & $5$ \\
\end{tabular}
\quad
\begin{tabular}{cccccccc}
$r_{32}$=$3$ & $6$ & $2$ & $4$ & $1$ & $7$ & $5$ & $0$ \\
$r_{33}$=$3$ & $2$ & $4$ & $1$ & $7$ & $5$ & $0$ & $6$ \\
$r_{34}$=$3$ & $4$ & $1$ & $7$ & $5$ & $0$ & $6$ & $2$ \\
$r_{35}$=$3$ & $1$ & $7$ & $5$ & $0$ & $6$ & $2$ & $4$ \\
$r_{36}$=$4$ & $2$ & $1$ & $6$ & $0$ & $7$ & $3$ & $5$ \\
$r_{37}$=$4$ & $1$ & $6$ & $0$ & $7$ & $3$ & $5$ & $2$ \\
$r_{38}$=$4$ & $6$ & $0$ & $7$ & $3$ & $5$ & $2$ & $1$ \\
$r_{39}$=$4$ & $0$ & $7$ & $3$ & $5$ & $2$ & $1$ & $6$ \\
$r_{40}$=$4$ & $7$ & $3$ & $5$ & $2$ & $1$ & $6$ & $0$ \\
$r_{41}$=$4$ & $3$ & $5$ & $2$ & $1$ & $6$ & $0$ & $7$ \\
$r_{42}$=$4$ & $5$ & $2$ & $1$ & $6$ & $0$ & $7$ & $3$ \\
$r_{43}$=$5$ & $6$ & $3$ & $2$ & $7$ & $0$ & $1$ & $4$ \\
$r_{44}$=$5$ & $3$ & $2$ & $7$ & $0$ & $1$ & $4$ & $6$ \\
$r_{45}$=$5$ & $2$ & $7$ & $0$ & $1$ & $4$ & $6$ & $3$ \\
$r_{46}$=$5$ & $7$ & $0$ & $1$ & $4$ & $6$ & $3$ & $2$ \\
$r_{47}$=$5$ & $0$ & $1$ & $4$ & $6$ & $3$ & $2$ & $7$ \\
\end{tabular}
\quad
\begin{tabular}{cccccccc}
$r_{48}$=$5$ & $1$ & $4$ & $6$ & $3$ & $2$ & $7$ & $0$ \\
$r_{49}$=$5$ & $4$ & $6$ & $3$ & $2$ & $7$ & $0$ & $1$ \\
$r_{50}$=$6$ & $5$ & $7$ & $4$ & $3$ & $1$ & $0$ & $2$ \\
$r_{51}$=$6$ & $7$ & $4$ & $3$ & $1$ & $0$ & $2$ & $5$ \\
$r_{52}$=$6$ & $4$ & $3$ & $1$ & $0$ & $2$ & $5$ & $7$ \\
$r_{53}$=$6$ & $3$ & $1$ & $0$ & $2$ & $5$ & $7$ & $4$ \\
$r_{54}$=$6$ & $1$ & $0$ & $2$ & $5$ & $7$ & $4$ & $3$ \\
$r_{55}$=$6$ & $0$ & $2$ & $5$ & $7$ & $4$ & $3$ & $1$ \\
$r_{56}$=$6$ & $2$ & $5$ & $7$ & $4$ & $3$ & $1$ & $0$ \\
$r_{57}$=$7$ & $3$ & $6$ & $1$ & $5$ & $4$ & $2$ & $0$ \\
$r_{58}$=$7$ & $6$ & $1$ & $5$ & $4$ & $2$ & $0$ & $3$ \\
$r_{59}$=$7$ & $1$ & $5$ & $4$ & $2$ & $0$ & $3$ & $6$ \\
$r_{60}$=$7$ & $5$ & $4$ & $2$ & $0$ & $3$ & $6$ & $1$ \\
$r_{61}$=$7$ & $4$ & $2$ & $0$ & $3$ & $6$ & $1$ & $5$ \\
$r_{62}$=$7$ & $2$ & $0$ & $3$ & $6$ & $1$ & $5$ & $4$ \\
$r_{63}$=$7$ & $0$ & $3$ & $6$ & $1$ & $5$ & $4$ & $2$ \\
\end{tabular}
\end{table}

\section{Optimal measurement order algorithm}
\label{appendix:algorithm}
As mentioned in the main text, we consider an adaptive algorithm that integrates dynamic programming, heuristic algorithms, and simulated annealing to optimize the sequence of measurement settings. The algorithm consists of two parts: a main program and the subfunctions required to implement the main program, as shown below.
\begin{algorithm}[H]
\caption{Qudit Measurement Sequence Optimization - Main Algorithm}
\label{alg:qudit_optimization_main}

\SetKwInOut{Input}{Input}
\SetKwInOut{Output}{Output}
\SetKwFunction{BuildCostMatrix}{BuildCostMatrix}
\SetKwFunction{HeldKarp}{HeldKarp}
\SetKwFunction{ClusterNN}{ClusterNearestNeighborV2}
\SetKwFunction{TwoOpt}{TwoOptOptimizationV2}
\SetKwFunction{SimulatedAnnealing}{SimulatedAnnealing}
\SetKwFunction{ParseConfig}{ParseMeasurementConfig}
\SetKwFunction{HammingDist}{ComputeHammingDistance}

\Input{Measurement configurations $\mathcal{C} = \{c_1, c_2, \ldots, c_n\}$, qudit dimension $d$, optimization method $m$}
\Output{Optimized sequence $\pi^*$ with minimum switching cost, total cost $C_{\text{total}}$}

\BlankLine

\tcp{Phase 1: Cost Matrix Construction with Caching}
$C \leftarrow \text{zero matrix of size } n \times n$\;
$\mathcal{P} \leftarrow \emptyset$ \tcp*{Set of parsed configurations}
\For{$i = 1$ \KwTo $n$}{
    $p_i \leftarrow$ \ParseConfig{$c_i$} \tcp*{Parse into tuple for hashing}
    $\mathcal{P} \leftarrow \mathcal{P} \cup \{p_i\}$\;
}
\For{$i = 1$ \KwTo $n$}{
    \For{$j = i+1$ \KwTo $n$}{
        $d_{ij} \leftarrow$ \HammingDist{$p_i, p_j$} \tcp*{Cached Hamming distance}
        $C[i,j] \leftarrow d_{ij}$, $C[j,i] \leftarrow d_{ij}$ \tcp*{Symmetric matrix}
    }
}

\BlankLine

\tcp{Phase 2: Adaptive Algorithm Selection}
\Switch{$m$}{
    \Case{\textnormal{'exact'}}{
        \tcp{Held-Karp with memory optimization}
        \textbf{Initialize} $dp$ as dictionary, $parent$ as dictionary\;
        \For{$i = 1$ \KwTo $n$}{
            $dp[\{i\}, i] \leftarrow 0$ \tcp*{Single-node paths}
        }
        \For{$k = 2$ \KwTo $n$}{
            \For{each subset $S$ with $|S| = k$}{
                \For{each $j \in S$}{
                    $prev \leftarrow S \setminus \{j\}$\;
                    $dp[S, j] \leftarrow \min_{i \in prev} \{dp[prev, i] + C[i,j]\}$\;
                    Store $parent[S, j] \leftarrow \arg\min_{i \in prev} \{dp[prev, i] + C[i,j]\}$\;
                }
            }
        }
        $\pi^* \leftarrow$ backtrack from $\min_j dp[\{1,\ldots,n\}, j]$\;
    }
    \Case{\textnormal{'heuristic'}}{
        \tcp{Enhanced clustering + 2-opt}
        $\pi_{\text{init}} \leftarrow$ \ClusterNN{$C, \mathcal{P}$} \tcp*{Improved clustering NN}
        $\pi^* \leftarrow$ \TwoOpt{$\pi_{\text{init}}, C$} \tcp*{Time-limited 2-opt}
    }
    \Case{\textnormal{'sa'}}{
        $\pi_{\text{init}} \leftarrow$ \ClusterNN{$C, \mathcal{P}$} \tcp*{Good initial solution}
        $\pi^* \leftarrow$ \SimulatedAnnealing{$C, \pi_{\text{init}}$} \tcp*{Simulated annealing}
    }
    \Case{\textnormal{'auto'}}{
        \eIf{$n \leq 12$}{
            Use exact method\;
        }{
            \eIf{$n \leq 50$}{
                Use heuristic method\;
            }{
                Use simulated annealing\;
            }
        }
    }
}

\BlankLine

\tcp{Phase 3: Performance Evaluation}
$C_{\text{total}} \leftarrow \sum_{i=1}^{n-1} C[\pi^*_i, \pi^*_{i+1}]$\;
$C_{\text{random}} \leftarrow$ average cost of random permutations\;
$\text{improvement} \leftarrow \frac{C_{\text{random}} - C_{\text{total}}}{C_{\text{random}}} \times 100\%$\;

\BlankLine
\Return{$\pi^*$, $C_{\text{total}}$, improvement}
\end{algorithm}

\begin{algorithm}[H]
\caption{Cluster-based Nearest Neighbor Algorithm}
\label{alg:cluster_nn}

\SetKwFunction{ClusterNN}{ClusterNearestNeighborV2}
\SetKwProg{Fn}{Function}{:}{}

\Fn{\ClusterNN{$C, \mathcal{P}$}}{
    \tcp{Feature-based clustering using basis type statistics}
    \For{$i = 1$ \KwTo $n$}{
        $\mathbf{f}_i \leftarrow [\text{count}_a(p_i), \text{count}_b(p_i), \text{count}_c(p_i)]$ \tcp*{Basis type features}
    }
    $\{\mathcal{K}_1, \ldots, \mathcal{K}_k\} \leftarrow$ K-means clustering of $\{\mathbf{f}_1, \ldots, \mathbf{f}_n\}$\;
    
    \tcp{Greedy cluster ordering by connectivity}
    $\text{order} \leftarrow [\arg\max_{j} |\mathcal{K}_j|]$ \tcp*{Start with largest cluster}
    $\text{remaining} \leftarrow \{1,\ldots,k\} \setminus \text{order}$\;
    \While{remaining $\neq \emptyset$}{
        $c_{\text{current}} \leftarrow$ all nodes in clusters currently in order\;
        $j^* \leftarrow \arg\min_{j \in \text{remaining}} \min_{u \in c_{\text{current}}, v \in \mathcal{K}_j} C[u,v]$\;
        Append $j^*$ to order, remove from remaining\;
    }
    
    \tcp{Nearest neighbor within each cluster}
    $\pi \leftarrow \emptyset$\;
    \For{$j \in \text{order}$}{
        \If{$\pi = \emptyset$}{
            $v \leftarrow$ first node in $\mathcal{K}_j$\;
        }\Else{
            $v \leftarrow \arg\min_{u \in \mathcal{K}_j} C[\pi_{\text{last}}, u]$\;
        }
        $\text{visited} \leftarrow \{v\}$, append $v$ to $\pi$\;
        \While{$\text{visited} \neq \mathcal{K}_j$}{
            $v \leftarrow \arg\min_{u \in \mathcal{K}_j \setminus \text{visited}} C[v, u]$\;
            Append $v$ to $\pi$, add $v$ to visited\;
        }
    }
    \Return{$\pi$}
}
\end{algorithm}

\begin{algorithm}[H]
\caption{2-Opt Local Optimization Algorithm}
\label{alg:two_opt}

\SetKwFunction{TwoOpt}{TwoOptOptimizationV2}
\SetKwProg{Fn}{Function}{:}{}

\Fn{\TwoOpt{$\pi, C$}}{
    $\pi_{\text{current}} \leftarrow \pi$, improved $\leftarrow$ \textbf{true}\;
    $\text{startTime} \leftarrow \text{currentTime}()$\;
    \While{improved \textbf{and} (\text{currentTime}() - \text{startTime}) $< t_{\text{max}}$}{
        improved $\leftarrow$ \textbf{false}\;
        \For{$i = 1$ \KwTo $|\pi|-2$}{
            \For{$j = i+2$ \KwTo $|\pi|$}{
                \tcp{Incremental cost calculation}
                \If{$j + 1 \leq |\pi|$}{
                    $\Delta \leftarrow (C[\pi_i,\pi_j] + C[\pi_{i+1},\pi_{j+1}]) - (C[\pi_i,\pi_{i+1}] + C[\pi_j,\pi_{j+1}])$\;
                }\Else{
                    $\Delta \leftarrow C[\pi_i,\pi_j] - C[\pi_i,\pi_{i+1}]$\;
                }
                \If{$\Delta < 0$}{
                    Reverse segment $\pi[i+1:j]$ in $\pi_{\text{current}}$\;
                    improved $\leftarrow$ \textbf{true}\;
                    \textbf{break}\;
                }
            }
            \If{improved}{\textbf{break}}
        }
    }
    \Return{$\pi_{\text{current}}$}
}
\end{algorithm}

\begin{algorithm}[H]
\caption{Simulated Annealing Algorithm}
\label{alg:simulated_annealing}

\SetKwFunction{SimulatedAnnealing}{SimulatedAnnealing}
\SetKwProg{Fn}{Function}{:}{}

\Fn{\SimulatedAnnealing{$C, \pi_{\text{init}}$}}{
    $T \leftarrow T_0$, $\pi_{\text{current}} \leftarrow \pi_{\text{init}}$\;
    $\pi_{\text{best}} \leftarrow \pi_{\text{init}}$, $E_{\text{best}} \leftarrow \text{cost}(\pi_{\text{init}})$\;
    \For{$iter = 1$ \KwTo $iter_{\text{max}}$}{
        \tcp{Generate neighbor solution}
        \If{random() $< 0.5$}{
            \tcp{2-opt move}
            $i, j \leftarrow$ random indices with $i < j-1$\;
            $\pi' \leftarrow \pi_{\text{current}}[1:i] + \text{reverse}(\pi_{\text{current}}[i+1:j]) + \pi_{\text{current}}[j+1:]$\;
        }\Else{
            \tcp{Swap move}
            $\pi' \leftarrow \pi_{\text{current}}$\;
            $i, j \leftarrow$ two distinct random indices\;
            Swap $\pi'[i]$ and $\pi'[j]$\;
        }
        
        \tcp{Metropolis acceptance criterion}
        $\Delta E \leftarrow \text{cost}(\pi') - \text{cost}(\pi_{\text{current}})$\;
        \If{$\Delta E < 0$ \textbf{or} random() $< \exp(-\Delta E/T)$}{
            $\pi_{\text{current}} \leftarrow \pi'$\;
        }
        
        \tcp{Update best solution}
        \If{$\text{cost}(\pi_{\text{current}}) < E_{\text{best}}$}{
            $\pi_{\text{best}} \leftarrow \pi_{\text{current}}$\;
            $E_{\text{best}} \leftarrow \text{cost}(\pi_{\text{current}})$\;
        }
        
        $T \leftarrow \alpha \cdot T$ \tcp*{Geometric cooling}
        \If{$T < T_{\text{min}}$}{\textbf{break}}
    }
    \Return{$\pi_{\text{best}}$}
}
\end{algorithm}

\end{document}